\documentclass[namedreferences]{solarphysics}
\usepackage[hyperref,optionalrh,solaromanenum,natbib]{spr-sola-addons} 
\usepackage{graphicx}                    
\usepackage{amssymb}                    
\usepackage[usenames]{color}             
\usepackage{subfig}
\usepackage{siunitx}

\begin{document}

\begin{article}

\begin{opening}

\title{Neutral Hydrogen and its Emission Lines in the Solar Corona}

\author[addressref=ias,corref,email={jean-claude.vial@ias.u-psud.fr}]{\inits{J.-C.}\fnm{Jean-Claude}~\lnm{Vial}}
\author[addressref=ias]{\inits{G.}\fnm{Martine}~\lnm{Chane-Yook}}

\address[id=ias]{Institut d'Astrophysique Spatiale, CNRS, Universit\'e
Paris-Sud, Universit\'e Paris-Saclay, B\^at. 121, 91405 Orsay cedex, France}

\runningauthor{J.-C. Vial {\it et al}.}
\runningtitle{Neutral Hydrogen and its emission lines in the solar corona}

\begin{abstract}
Since the L$\alpha$ rocket observations of (Gabriel, {\it Solar Phys.} {\bf 21}, 392, 1971), it has been realized that the hydrogen (H) lines could be observed in the corona and offer an interesting diagnostic for the temperature, density, and radial velocity of the coronal plasma. Moreover, various space missions have been proposed to measure the coronal magnetic and velocity fields through polarimetry in H lines. A necessary condition for such measurements is to benefit from a sufficient signal-to-noise ratio. The aim of this article is to evaluate the emission in three representative lines of H for three different coronal structures. The computations have been performed with a full non-local thermodynamic-equilibrium (non-LTE) code and its simplified version without radiative transfer. Since all collisionnal and radiative quantities (including incident ionizing and exciting radiation) are taken into account, the ionization is treated exactly. Profiles are presented at two heights (1.05 and 1.9 solar radii, from Sun center) in the corona, and the integrated intensities are computed at heights up to five solar radii. We compare our results with previous computations and observations ({\it e.g.} L$\alpha$ from UVCS) and find a rough (model-dependent) agreement. Since the H$\alpha$ line is a possible candidate for ground-based polarimetry, we show that in order to detect its emission in various coronal structures, it is necessary to use a very narrow (less than 2~\AA~wide) bandpass filter.

\end{abstract}



\end{opening}

%
%
\section{Introduction}
It came as a surprise to discover that a hot and diluted medium such as the corona was emitting the ``cool'' L$\alpha$ line \citep{1971SoPh...21..392G}. Since this rocket observation (eclipse), many more L$\alpha$ observations have been performed ({\it e.g.} \citet{1994ESASP.373..363H} and \citet{1995SSRv...72...29K} by the \textit{Ultraviolet Coronal Spectrometer} (UVCS) on board \textit{Spartan}, and later on by the \textit{Ultraviolet Coronal Spectrometer} (UVCS) on board the \textit{Solar and Heliospheric Observatory} (SOHO) (\citet{1999A&A...342..592Z} and \citet{1999SSRv...87..265M})) with the help of a coronagraph. The emission was quickly identified as the resonance scattering of the chromospheric L$\alpha$ radiation by ``trace'' neutral hydrogen (usually taken as about $10^{-6}$ electron density). Because L$\alpha$ is the strong resonance line of the most abundant element (hydrogen) and because the emitted chromospheric profile is about as wide as the absorption profile of coronal neutral hydrogen, in the absence of velocity field, the emission is rather strong (about $10^{-6}$ the disk value at 1.5 R$_{\odot}$) and even stronger (a few $10^{-5}$) in a streamer \citep[\textit{e.g.}][]{1999ESASP.448.1193M}. This is why a major future mission, {\it Solar Orbiter}, includes the L$\alpha$ coronograph called \textit{Multi Element Telescope for Imaging and Spectroscopy} (METIS) \citep{2012SPIE.8443E..09A}. Moreover, it has been shown as early as 1982 \citep{1982SoPh...78..157B} that the line was sensitive to the Hanle effect (generally effective in weak magnetic fields). Consequently, many (space) projects have proposed polarimetric measurements in this line ({\it e.g.} the \textit{Small Explorer for Solar Eruptions} (SMESE) mission \citep{2007AdSpR..40.1787V} and more recently the \textit{Coronal UV spectro-polarimeter} (CUSP) on board the \textit{Solar magnetism eXplorer} (SolmeX) \citep{2012ExA....33..271P}, the \textit{MAGnetic Imaging Coronagraph} (MAGIC) on board the \textit{INvestigation of Solar-Terrestrial Activity aNd Transients} (INSTANT) mission (Lavraud {\it{et al.}} (2015), proposal to ESA), the \textit{MAGnetic Imaging Coronagraph} (MAGIC) on board the \textit{Magnetic Activity of the Solar Corona} (MASC) mission (Auch\`ere {\it{et al.}} (2015), proposal to ESA). It has also been proposed to use the L$\beta$ line for performing polarimetric measurements in the faint corona \citep{2012ExA....33..271P}.\\
However, until now, these projects are still at the proposal level for various reasons including the (relative) complexity of the instrumentation and the fact that polarimetry is ``photon-hungry'' and requires large apertures in the ultra-violet (UV). Apart from radio observations above active regions, another path towards coronal polarimetry has been pursued with ground-based infrared (IR) observations measuring the Zeeman effect \citep{2000ApJ...541L..83L,2008SoPh..247..411T}.\\
Another possibility has been opened with eclipse polarization measurements in red and green channels by \citet{2013Ge&Ae..53..901K} from which these authors concluded that ``the polarization excess (green-red) can be explained by the presence of neutral hydrogen in the corona'' \citep[see also][]{2014A&A...567A...9D}. As mentioned by \citet{2013Ge&Ae..53..901K}, \citet{1976ApJ...209..927P} and \citet{2011A&A...530L...1M} had already concluded that H$\alpha$ contributed to coronagraphic images of transients and coronal mass ejections (CMEs), respectively. Actually this was demonstrated with eclipse measurements made at the Canada-France-Hawa\"{\i} Telescope (CFHT) in 1991 where \citet{1992ESASP.344...87V} detected a plasmoid in the solar corona that \citet{1994A&A...281..249K} interpreted as emitting essentially in the H$\alpha$ line, a claim discussed later on by \citet{2000A&A...353..786Z} who proposed an upper limit of the emission of about 2 \% of the background corona. Moreover, the Hanle effect has been successfully used in the H$\alpha$ line, in spite of its optical thickness, in magnetic-field measurements in prominences \citep[\textit{e.g.}][]{1981SoPh...71..285L,1981A&A...100..231B}.\\
In order to derive the plasma properties (including magnetic field) it is imperative to take into account all processes involved and first of all the proper ionization degree. The aim of this work is to compute most observable parameters (line profile, intensity) for the main H lines exactly emitted by (three) typical regions of the solar corona and at various heights above the limb. \\
In Section \ref{section2}, we present the full non-LTE computations derived from 1D non-LTE codes which are adapted to the geometry of the corona. In Section \ref{section3}, we focus on profiles obtained in the L$\alpha$, L$\beta$ and H$\alpha$ lines at various heights. In Section \ref{comparaison_Ly_obs}, we also compare with other computational and observational results in the L$\alpha$ and L$\beta$ lines. In Section \ref{section4}, we provide the variation with altitude of the ionization degree for the three models. In Section \ref{section5}, we discuss all our results, and we pay some attention to the possibility of observing the H$\alpha$ line in the corona. In Section \ref{section6}, we conclude on further improvements in the computations. \\
We also present in the Appendix an approximation to obtain the H$\alpha$ intensity from the (measured) L$\alpha$ emission. \\

\section{Non-LTE Computations}\label{section2}
As far as hydrogen lines are concerned, we remark that the radiative output of structures such as prominences and the corona is dominated by the incident radiation and the resonance scattering that follows. The L$\alpha$ case has been extensively studied since the pioneering work of \citet{1976ApJ...205..273H} who built a non-LTE code adapted to one dimensional cool layers located in the corona illuminated by the photospheric, chromospheric and coronal radiation. Much modeling has been performed since \citep[\textit{e.g.}][]{1993A&AS...99..513G} where the prominence was considered as an isothermal and isobaric layer. More recently, prominence--corona transition regions (PCTR) have been grafted to the homogeneous layer on the basis of magnetohydrostatic (mhs) equilibrium by \citet{1999A&A...349..974A} and then consistent mhs models of threads have been combined with the addition of inhomogeneous layers \citep{2005A&A...442..331H}; for comprehensive reviews of this modeling, see \citet{2010SSRv..151..243L}, \citet{2014IAUS..300...59G}, \citet{2015ASSL..415..131L} and \citet{2015ASSL..415..103H}. 
In all these recent models, there is basically a cool core (T $\approx$ 10 000 K) and a PCTR where physical quantities vary drastically up to the corona itself where the density is low enough for all the lines and continua of all elements (including hydrogen) to be optically thin. In the low corona, the Lyman lines have been treated by \citet{1987ApJ...315..706N} with the simple assumption of resonance scattering to begin with and later on with the inclusion of electron collisions, which allows for more information to be drawn from L$\alpha$ and L$\beta$ observations \citep{1982SSRv...33...17W,1997SoPh..175..645R}. As shown by \citet{2006A&A...455..719L} in the case of a coronal streamer, the L$\alpha$ line is essentially radiatively formed but the L$\beta$ line becomes increasingly collisionnally driven with increasing altitude. Moreover, as discussed in Section \ref{section4}, the ionization has been computed up to now within the simplified model of \citet{1971SoPh...21..392G}. \\
For various reasons (including its non-detection even with the best coronagraphs, the best seeing, and the most narrow filters), the coronal H$\alpha$ line has never been studied in detail. It is formed in the corona through various processes (radiative and partly collisional) involving at least the three first levels and possible cascades from higher levels including the continuum. In the Appendix, we propose a simple approach for computing H$\alpha$, an approach which involves radiative processes and the first three levels only.\\
In order to perform an exact computation of the hydrogen lines (which implies to compute a high number of population levels, including the continuum, with all processes taken into account), we decided to follow a ``non-LTE radiative transfer'' approach although it is clear that because of the low neutral density in the corona (about $10^{-6}$ times the electron density) all hydrogen lines are optically thin and all resonance scatterings are not followed by a second scattering. The advantage is that the bulk of the code is available and well controlled, in such a way that it is then possible to treat many hydrogen levels (including the continuum) with precise incident radiation altitude-dependent profiles which are critical for the population of H levels. This approach does not simplify the incident profiles, contrary to \citet{2005ApJ...622..737A} or \citet{2015A&A...577A..34D}, who use a combination of three gaussians. It formally allows for the consideration of incident profiles varying with distance to the limb, or with the presence of close active regions, or the variation with activity. As far as the thermodynamic parameters, density [$n_e$] and temperature [$T_e$] of the corona are concerned, they vary with radial distance (see Section \ref{sect_2.2}). As already mentioned, such an approach has been followed by \citet{2006A&A...455..719L} who computed the L$\alpha$ and L$\beta$ integrated intensities in a streamer defined through the three-fluid solar-wind model of \citet{2006JGRA..111.8106L}. However, we proceeded with a different method and different models as shown below.
\subsection{The Computational Method}
The initial prominence code, PROM5 (\textsf{https://idoc.ias.u-psud.fr/ \\MEDOC/Radiative transfer codes/PROM5}) is modified as follows. Instead of a cool core with an extended PCTR, we simply have no layer with low temperature but only a so-called PCTR with actual coronal conditions in a spherically symmetric configuration where the density decreases from the impact parameter position toward both sides of the line-of-sight (LOS). The medium being optically thin in all lines and in all directions, the incident radiation is exactly computed at each radial distance. Actually, the computation of the L$\alpha$ line emitted when the line-of-sight crosses the limb (and where a plane-parallel LOS computation makes no sense) has been performed in the more difficult case of an optically thick atmosphere \citep{1970IAUS...36..260V}. In order to simplify the computations, we adopted the same model extension, for all LOS located at altitudes between 1.05 and 5 R$_{\odot}$. This means that the boundaries along each LOS are (symmetrically) defined by the last external layer of the radial model. Through extrapolation, we checked that the neglected (out-of-model) parts along the LOS did not contribute much to the computed opacities ({\it i.e.} intensities): for L$\alpha$ we found less that 1 \% at 1.05 R$_{\odot}$ and much less for the other lines and lower altitudes. For higher altitudes ({\it i.e.} 5 R$_{\odot}$), the neglected opacity reaches 4 \% for the coronal hole model.\\
The dedicated coronal code, PROMCOR is available at:\\
\textsf{https://idoc.ias.u-psud.fr/MEDOC/Radiative transfer codes/PROMCOR} \\
where a typical coronal hole model is proposed as an input.

\subsection{The Physical Models}\label{sect_2.2}
\begin{figure}[H]
\centering
\includegraphics[width=0.9\textwidth,keepaspectratio]{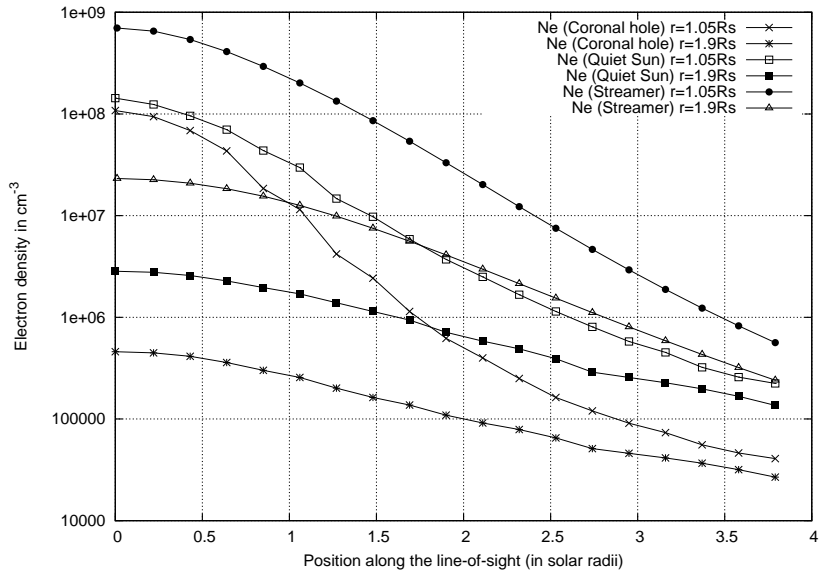}
\caption{Electron densities for the quiet Sun, coronal hole, and streamer models at two altitudes: 1.05 and 1.9 R$_{\odot}$, along half of the LOS.}\label{fig1}
\end{figure}

We limited our study to three different models: the quiet corona and polar coronal hole models provided by \citet{1977asqu.book.....A} and the streamer model of \citet{2014ApJ...781..100G}. The streamer model is provided in 2D as an angular diverging slab with a Gaussian shape with a constant temperature of $1.43\times 10^6$ K as derived from the measurements from the \textit{Sun Watcher using Active Pixel System Detector and Image Processing} (SWAP) and the \textit{EUV imaging spectrometer} (EIS) on board \textit{Hinode}. We simplified the model into a spherically symmetric one with a temperature of $10^6$ K in order to better compare with the quiet corona values of \citet{1977asqu.book.....A}. Note that \citet{2015A&A...577A..34D} found a temperature range of $5\times 10^5$ K\,--\,$1.5\times 10^6$ K between 2 and 6 R$_{\odot}$, over a large range of position angles. We adopted the radial density provided by Equation (16) and Table 2 of \citet{2014ApJ...781..100G}. In doing so, we do not have to take into account any background contribution since we have a unique geometrical model but we are aware that all our LOS intensities are overestimated as compared to the actual Goryaev model. In order to roughly compare the L$\alpha$ and L$\beta$ results with the streamer model we selected two lines of sight: one at 1.05 R$_{\odot}$ and the other at 1.9 R$_{\odot}$. \\
The initial density profile of the three models (quiet Sun, coronal hole, and streamer) along the two halfs of the LOS is shown in Figure \ref{fig1}. \\
We can note that our streamer electron density (half) profiles along the LOS have roughly the same magnitudes and shapes as the streamer profile of \citet{2006A&A...455..719L}.\\
As far as the temperature profile is concerned, as mentioned above, we used the temperature variation with altitude of \citet{1977asqu.book.....A} for the quiet Sun; we took a constant temperature of 800,000 K \citep[see][]{1998A&A...336L..90D} for the coronal hole and, as mentioned above, kept a constant temperature of $10^6$ K for the streamer model \citep[see][]{2014ApJ...781..100G}.
 
\section{Computed Profiles and Variation of the Integrated Intensities with the Altitude of the LOS}\label{section3}
\begin{figure}[H]
\centering
\includegraphics[width=0.9\textwidth,keepaspectratio]{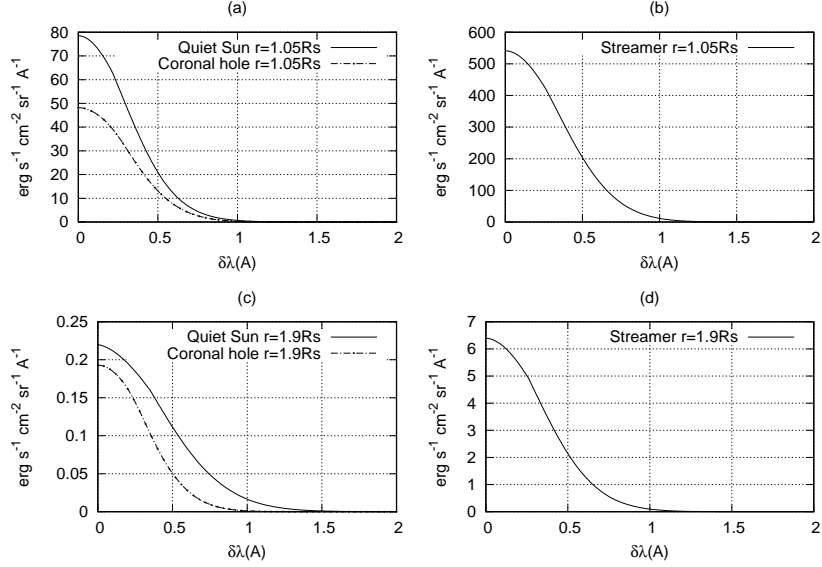}
\caption{Half-profiles of the L$\alpha$ line for LOS at 1.05 (first row, (a) and (b)) and at 1.9 R$_{\odot}$ (second row, (c) and (d)). \\On the left column ((a) and (c)) the half-profiles for the quiet-Sun and coronal-hole models. \\On the right column ((b) and (d)) the half-profiles for the streamer model.}\label{fig2}
\end{figure}

\begin{figure}[H]
\centering
\includegraphics[width=0.9\textwidth,keepaspectratio]{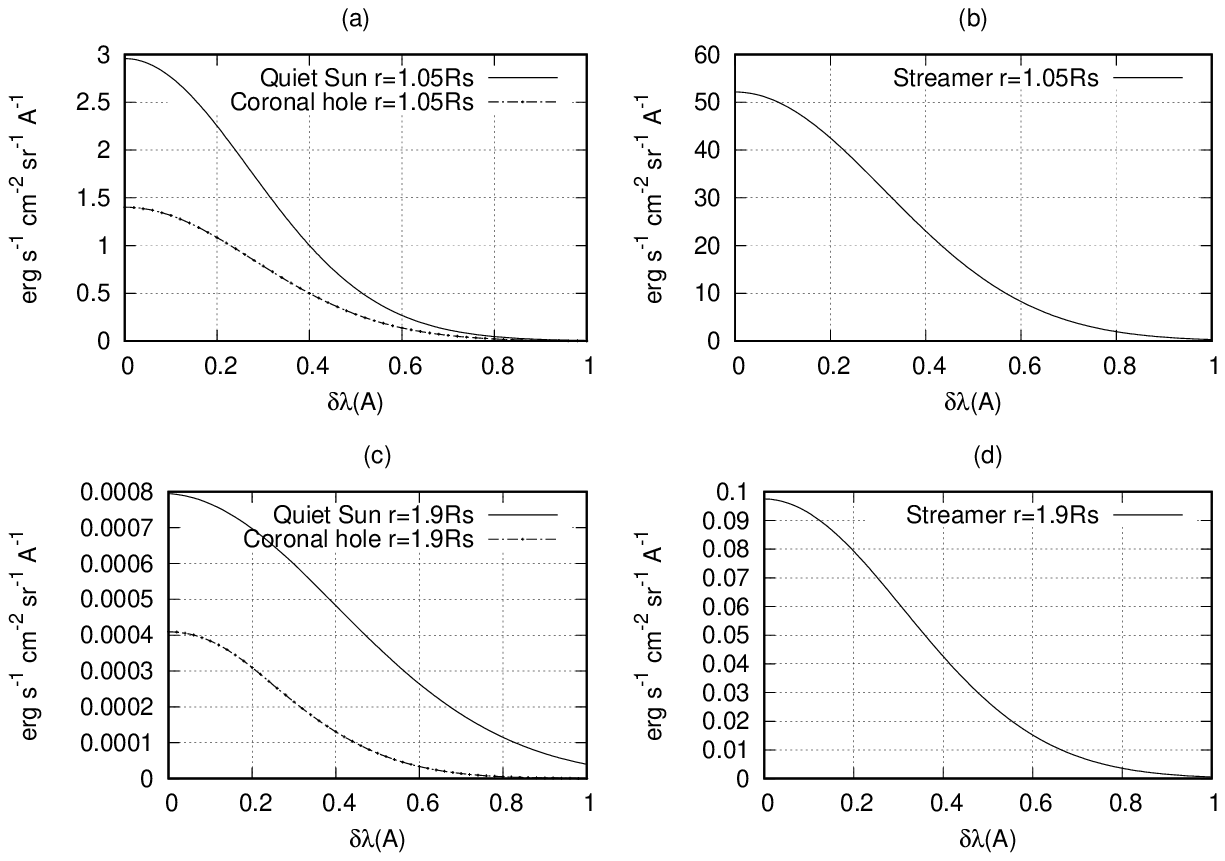}
\caption{Half-profiles of the L$\beta$ line for LOS at 1.05 (first row, (a) and (b)) and at 1.9 R$_{\odot}$ (second row, (c) and (d)). \\On the left column ((a) and (c)) the half-profiles for the quiet-Sun and coronal-hole models. \\On the right column ((b) and (d)) the half-profiles for the streamer model.}\label{fig3}
\end{figure}

\begin{figure}[H]
\centering
\includegraphics[width=0.9\textwidth,keepaspectratio]{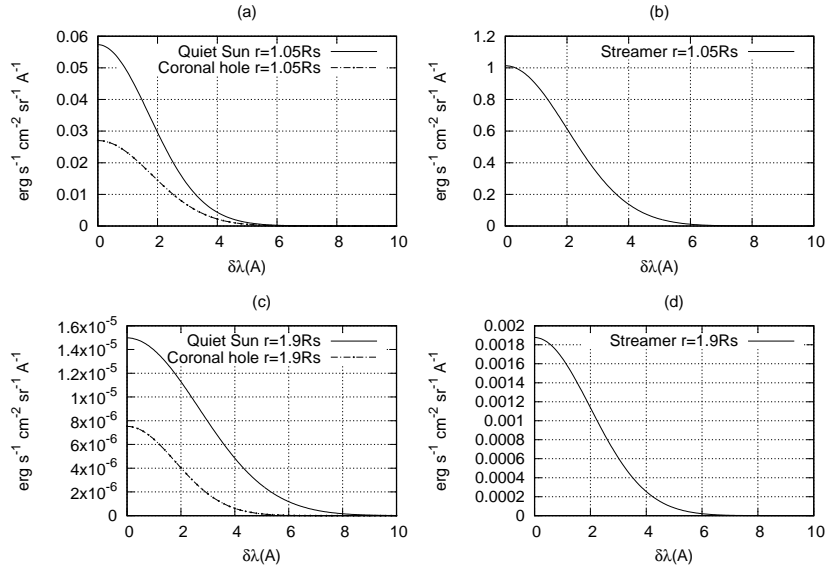}
\caption{Half-profiles of the H$\alpha$ line for LOS at 1.05 (first row, (a) and (b)) and at 1.9 R$_{\odot}$ (second row, (c) and (d)). \\On the left column ((a) and (c)) the half-profiles for the quiet-Sun and coronal-hole models. \\On the right column ((b) and (d)) the half-profiles for the streamer model.}\label{fig4}
\end{figure}

\begin{figure}[H]
\centering
\includegraphics[width=.9\textwidth,keepaspectratio]{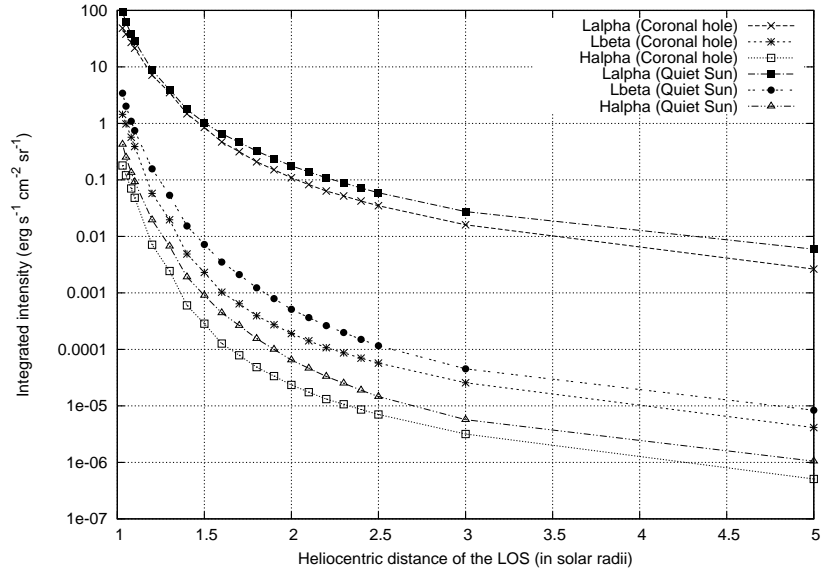} 
 \caption{Variation of the integrated intensities of the L$\alpha$, L$\beta$, and H$\alpha$ lines [in erg~s$^{-1}$cm$^{-2}$sr$^{-1}$] with the position of the LOS (in solar radii) for the quiet-Sun and coronal-hole model where continuum absorption is omitted.}\label{fig5-1} 
\end{figure}

\begin{figure}[H]
\centering
\includegraphics[width=.9\textwidth,keepaspectratio]{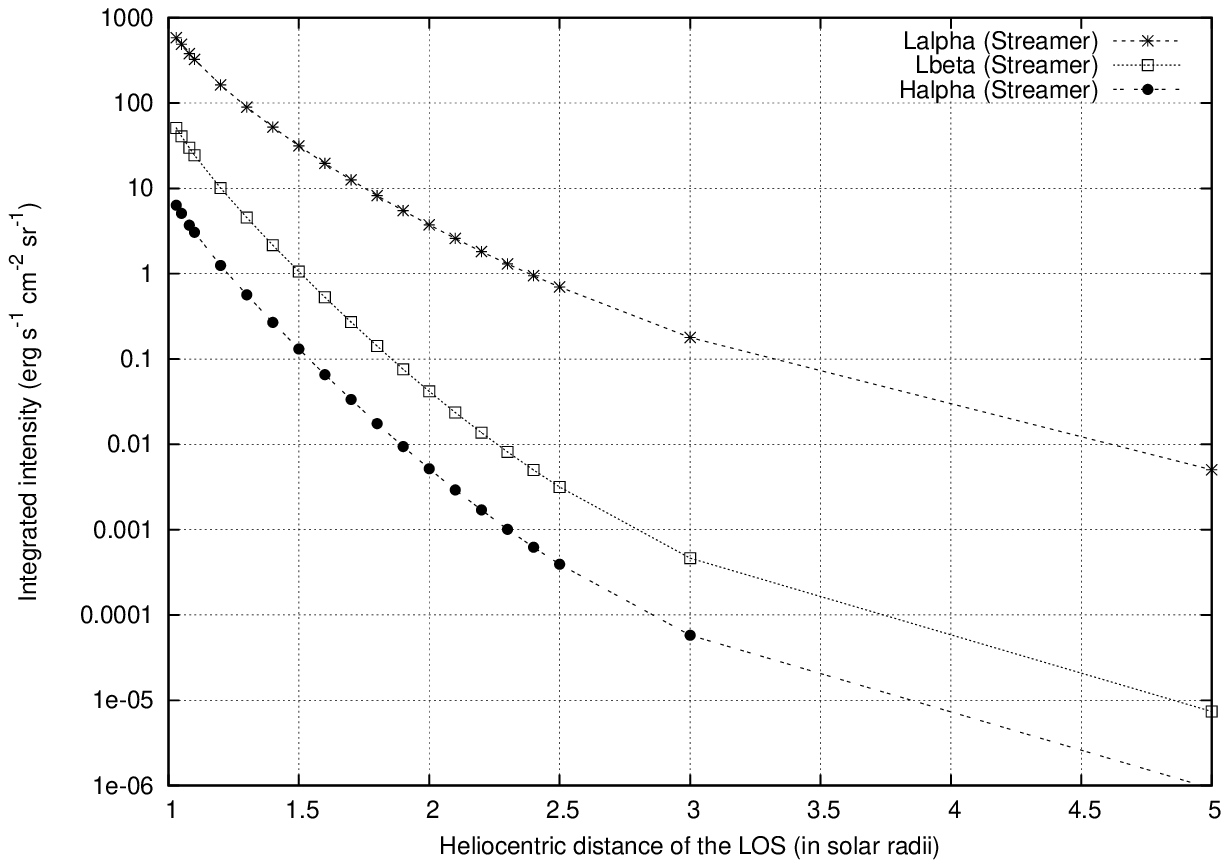} 
\caption{Variation of the integrated intensities of the L$\alpha$, L$\beta$, and H$\alpha$ lines [in erg~s$^{-1}$cm$^{-2}$sr$^{-1}$] with the position of the LOS (in solar radii) for the streamer model where continuum absorption is omitted.}\label{fig5-2}
\end{figure}

Since we assume no velocity field (neither radial nor along the LOS), we only compute half-profiles which are shown in Figures \ref{fig2} (L$\alpha$), \ref{fig3} (L$\beta$), and \ref{fig4} (H$\alpha$). Note that we limit ourselves to the presentation of the three lines but more lines have been computed since we have a five-level atom.\\
It is important to stress that the H$\alpha$ (half) profiles in Figure \ref{fig4} have been computed without any continuum absorption. This allows to compare the integrated intensities in the three main lines on one hand (Figures \ref{fig5-1} and \ref{fig5-2}) and, as far as the streamer is concerned, to compare with the L$\alpha$, L$\beta$ results of \citet{2006A&A...455..719L}, on the other hand. However, as shown in this section, the inclusion of absorption is critical for the H$\alpha$ line.\\
First, we note that for all models (Figures \ref{fig5-1} and \ref{fig5-2}) the L$\beta$ intensity is an order of magnitude smaller than the L$\alpha$ intensity at $r=1.05~R_\odot$, leading to a ratio (0.07) which is higher than the ratio of the incident intensities (between 0.011 and 0.014 according to \citet{2012A&A...542L..25L}). This ratio becomes $1.3\times 10^{-2}$ at 1.9 R$_{\odot}$, a value equal to the ratio of the incident intensities and to be compared with the $2\times 10^{-3}$ observed value of \citet{2003ApJ...597.1118C} at a distance of 2.3 R$_{\odot}$ in a pre-CME streamer (and a value slightly higher in the CME, or its ``prominence core'', as stated by these authors). This can be interpreted as the result of the large (average) densities in our one-dimensional streamer model, which increase the collisional component of the L$\beta$ intensity and the excitation of level 3 through 1--2 and 2--3 absorption and spontaneous emission from level 3 to level 1. 

\begin{figure}[H]
\centering
\includegraphics[width=1\textwidth,keepaspectratio]{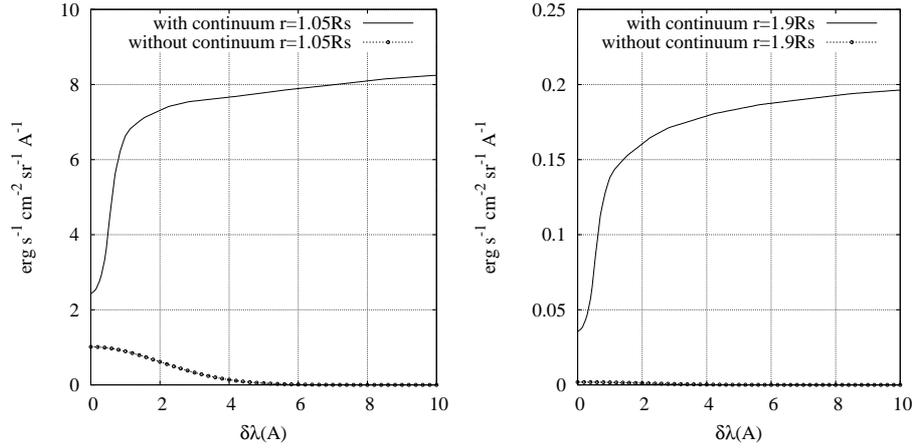}
\caption{Comparison of H$\alpha$ half profiles with and without continuum for the streamer model. Left: at R=1.05 R$_{\odot}$. Right: at 1.9 R$_{\odot}$.} \label{fig6-1}
\end{figure}

\begin{figure}[H]
\centering
\includegraphics[width=1\textwidth,keepaspectratio]{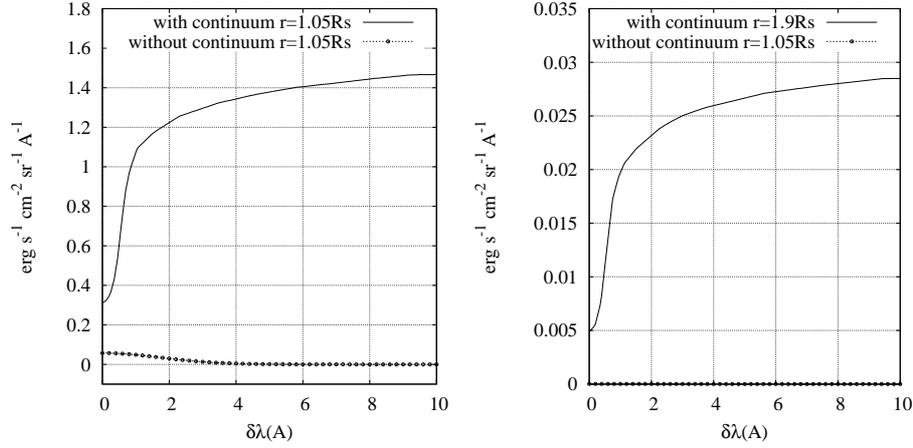}
\caption{Comparison of H$\alpha$ half profiles with and without continuum for the quiet-Sun model. Left: at R=1.05 R$_{\odot}$. Right: at 1.9 R$_{\odot}$.} \label{fig6-2}
\end{figure}

\begin{figure}[H]
\centering
\includegraphics[width=1\textwidth,keepaspectratio]{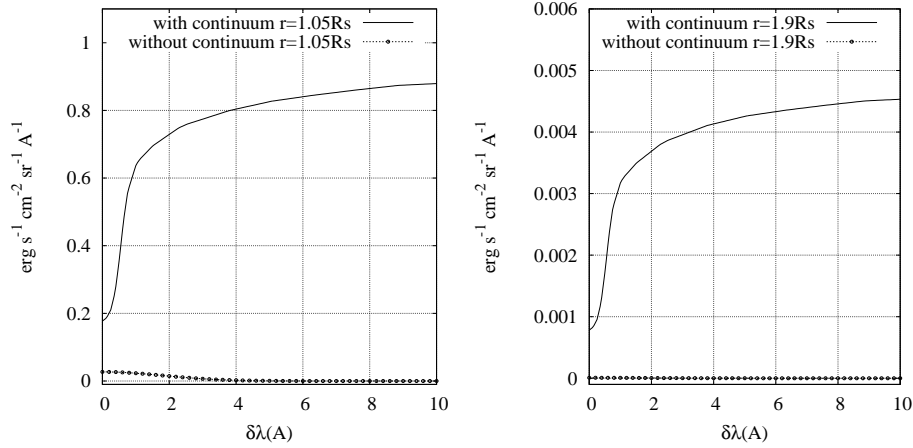}
\caption{Comparison of H$\alpha$ half profiles with and without continuum for the coronal-hole model. Left: at R=1.05 R$_{\odot}$. Right: at 1.9 R$_{\odot}$.} \label{fig6-3}
\end{figure}

Second, another interesting feature is the ratio between the L$\beta$ and H$\alpha$ integrated intensities which appears to be rather constant with the altitude and the model. Whatever the model and the altitude (see Figures \ref{fig5-1} and \ref{fig5-2}), the value found is about eight which transforms into 1.2 when we compute the ratio of intensities in number of photons instead of energy intensities. We can compare this value with the ratio of spontaneous emission factors [$A_{ij}$] in the L$\beta$ and L$\alpha$ lines (Heinzel, 2016, private communication). This is made difficult by the fact that the L$\beta$ line has three components and the H$\alpha$ line has eight components \citep{2015Kramida}. In our computations, we adopted $A_{31} = 5.5\times 10^7$ and $A_{32} = 4.4\times 10^7$ which leads to $A_{31}/A_{32} = 1.25$, a value not very different from the ratio of the number of emitted photons. This is no surprise since the two lines share the same upper level and all emission processes other than spontaneous emission are negligible. 
\newline
Third, our code can take into account the various continuum contributions (Thomson scattering, $\mathrm{H}^{-}$, Rayleigh, etc) but the only significant contribution comes from Thomson scattering, since the plasma is quite fully ionized. With Thomson opacity included (about $3\times 10^{-6}$) the intensity in the wings of H$\alpha$ is notably increased (Figures \ref{fig6-1}, \ref{fig6-2} and \ref{fig6-3}) and the profile is now in absorption. The relative absorption can be easily compared with the continuum within an order of magnitude. Even if the Thomson cross-section is about 11 orders of magnitude smaller than the H$\alpha$ absorption cross-section, we must take into account that the ratio of neutral (level 1) hydrogen to electron density populations is about $10^{-6}$ -- $10^{-7}$ (see Section \ref{section4}) or lower, as we shall see in section \ref{section4}, at the location of the impact parameter ({\it i.e.} at the closest position to the Sun on the LOS) for the Allen model (see Figure \ref{fig7}). We also take into account that the ratio of level 2 to level 1 populations is less than $10^{-7}$ for the same model and at the same altitudes. \\
Consequently the H$\alpha$ contribution is about two orders of magnitude smaller than the Thomson scattering and it decreases with increasing altitude. The profile is in absorption (due to the fact that the H$\alpha$ incident profile is in absorption) and the depth of the depression is about 80 \% of the local continuum. The width of the absorption line is about 2~\AA. \\
We do not present the comparison of L$\alpha$ and L$\beta$ profiles with and without continuum absorption because the continuum contribution is found to be negligible. This confirms why the measurement of the Thomson scattering at the L$\alpha$ wavelength with UVCS has been very difficult and consequently the Thomson scattering does not impede the purely L$\alpha$ scattering \citep{2007AGUFMSH21A0298K}.

\section{Comparison with Lyman Observations}\label{comparaison_Ly_obs}

As far as L$\alpha$ is concerned, we compare with profiles and intensities of \citet{1982SSRv...33...17W}. These authors show the relative variation of L$\alpha$ with altitude (their Figure 2). They find a L$\alpha$ width (1/e halfwidth) which decreases from about 0.75 to 0.58~\AA~between 1.5 and 3.5 R$_{\odot}$. At 1.9 R$_{\odot}$, they find 0.7~\AA,~to be compared with our 0.65~\AA~(it should be recalled that their spectroscopic data are corrected from an instrumental width of 0.34~\AA).\\
As far as intensities are concerned, the Withbroe's \textit{et al.} values, when converted in cgs units, are higher than ours by about a factor three, at about all altitudes. This means that the ratio of intensities at 1.9 and 3.5 R$_{\odot}$ is the same (about ten) for measured and computed values.\\
As far as coronal holes are concerned, a UVCS/SOHO intensity varying between $4\times 10^9$ and $6\times 10^9$ photons~s$^{-1}$~cm$^{-2}$~sr$^{-1}$ is found by \citet{2008ApJ...677L.137B} (see their Figure 2) from the center to the far edge of a polar coronal hole at an altitude of 2.14 R$_{\odot}$. Once converted in erg~s$^{-1}$cm$^{-2}$sr$^{-1}$, these values (0.06 and 0.096) are quite comparable with our value (0.1 erg~s$^{-1}$cm$^{-2}$sr$^{-1}$) at the same altitude.\\ 
We also compare our results with streamer observations. As far as profiles are concerned, \citet{1999ESASP.448.1193M} find a 1/e halfwidth varying from 0.75~\AA~(their northern streamer) and 0.65~\AA~(their southern streamer), at an altitude between 1.6 and 2.6 R$_{\odot}$. We find (Figure \ref{fig2}) 0.5~\AA~at 1.9 R$_{\odot}$.\\
We also compare with \citet{2015A&A...577A..34D} who derived the H kinetic temperature (their Figure 11) from the L$\alpha$ half-width. Their values are higher than the one we adopted but the authors mentioned that their values are ``higher than those determined by \citet{2007A&A...475..707S} and \citet{2008A&A...488..303S}''. As far as intensities are concerned, \citet{1999ESASP.448.1193M} find 0.8 to 1.3 erg~s$^{-1}$cm$^{-2}$sr$^{-1}$ at 1.9 R$_{\odot}$ (their Figure \ref{fig2}) and \citet{2015A&A...577A..34D} find 0.9 and 0.01 erg~s$^{-1}$cm$^{-2}$sr$^{-1}$ at 1.9 and 5 R$_{\odot}$, respectively. Our values are very close: 0.7 and 0.01 erg~s$^{-1}$cm$^{-2}$sr$^{-1}$ at 1.9 and 5 R$_{\odot}$ respectively.

\section{Ionization Degree in the Low Corona}\label{section4}
\begin{figure}[H]
\centering
\includegraphics[width=0.9\textwidth,keepaspectratio]{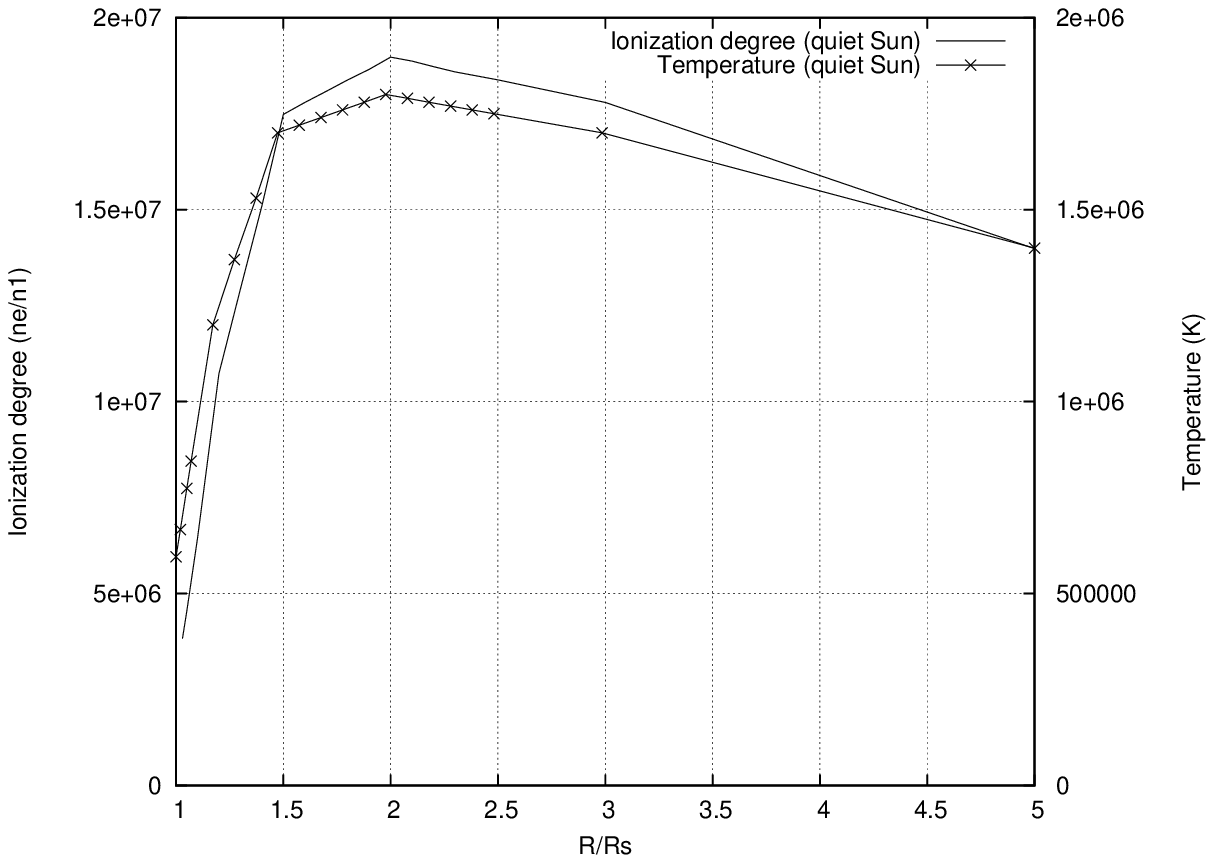}
\caption{Variation of the ionization degree (as defined by the ratio of electron to neutral hydrogen densities) and the temperature [K] with altitude (in solar radii) for the quiet-Sun model of \citet{1977asqu.book.....A}. We indicate the ionization degree with a solid line and the temperature with x signs.}
\label{fig7}
\end{figure}

For the three models considered, we are able to compute exactly the variation of the ionization, defined here as $n_e/n_{H_0}$ or $n_e/n_1$ where $n_{H_0}$ is the neutral hydrogen density and $n_1$ is the fundamental level population (we do not include He contribution to $n_e$). The values are in the expected range: for the quiet-Sun model, the ionization degree increases from $4\times 10^6$ close to the surface to $2\times 10^7$ between 1.5 and 2.5 R$_{\odot}$. Then it slightly decreases to $1.4\times 10^7$ at 5 R$_{\odot}$. This behaviour exactly matches the variation of temperature with altitude (Figure \ref{fig7}): with increasing temperature, all factors implying ionization, especially the collisional terms, increase \citep[see][]{1999ApJ...511..481C}. The values imply a $n_1/n_e$ ratio varying from about $2.5\times 10^{-7}$ to $5\times 10^{-8}$, which are lower, by an order of magnitude, than the values usually adopted in the corona \citep[{\textit{e.g.}}][]{1971SoPh...21..392G}. Note that Gabriel points out that his computation neglects photoionisation from level 2s. Here we stress that our computations include all hydrogen radiative terms. Note that the ionization degree is nearly constant for the coronal-hole ($6\times 10^6$) and streamer ($8\times 10^6$) models where the temperature is taken as constant (800,000 and $10^6$ K respectively).

\section{Discussion}\label{section5}
Let us note that since we have a five-level atom, we actually treat more transitions [L$\gamma$, H$\beta$, P$\alpha$, H$\gamma$, P$\beta$, H$\delta$, H$\epsilon$] than the three presented above. Half-profiles and integrated intensities of all these lines are available on the \textit{Multi Experiment Data and Operation Center} (MEDOC) site \\
\textsf{https://idoc.ias.u-psud.fr/MEDOC/Radiative transfer codes/PROMCOR}\\
We now raise the issue of the visibility of the hydrogen lines in the corona and in particular the polarimetry, since, as mentioned in many proposals and road maps \citep[see, {\it e.g.},][]{2015AdSpR..55.2745S} the measurement of the coronal magnetic field, is now a major objective in solar physics.\\
First, as shown in Section 4, we are not surprised as far as the L$\alpha$ line is concerned since the UVCS measurements (Figure \ref{fig8}) at a distance as low as 1.5 R$_{\odot}$ are close to our computations for our streamer model (see Figure \ref{fig5-2}). The L$\alpha$ variation of Figure \ref{fig5-2} shows that at 2.5 R$_{\odot}$ (1.5 R$_{\odot}$ above the surface) the L$\alpha$ intensity is about $1.4\times 10^{-5}$ the disk intensity. This altitude is the maximum altitude where linear polarization measurements can provide useful information about the coronal magnetic field through the Hanle effect \citep{2010A&A...511A...7D}. This value seems to be compatible with the scattering performances of the proposed instrumentation \citep[\textit{e.g.}][]{2007AdSpR..40.1787V}. However, it is clear that the coronal intensity is about a magnitude lower in a quiet corona and still lower in coronal holes (Figure \ref{fig5-1}) which means that measurements will be challenging at those locations.\\
Second, as for L$\beta$, in our streamer model the intensity variation with distance compares relatively well to the results of \citet{2006A&A...455..719L} in terms of number of photons. According to \citet{2013JGRA..118..967G}, the L$\beta$ intensity is 0.03 erg~s$^{-1}$cm$^{-2}$sr$^{-1}$ at 1.9 R$_{\odot}$ where we find 0.09 (note that the L$\alpha$ values are closer: 8 for our computations and 6.4 for \citet{2013JGRA..118..967G} at 1.9 R$_{\odot}$). The ratio L$\beta$/L$\alpha$ is about $10^{-3}$ as in \citet{2006A&A...455..719L} at 3 R$_{\odot}$ but it is about $10^{-1}$ instead of $10^{-2}$ in \citet{2006A&A...455..719L} for 1 R$_{\odot}$. With the quiet-Sun (Allen) model, the L$\beta$/L$\alpha$ ratio is lower than $10^{-1}$ at 1 R$_{\odot}$ and $2\times 10^{-3}$ at 2.5 R$_{\odot}$, which means that polarization measurements in the L$\beta$ line \citep{2012ExA....33..271P} will face serious difficulties because of the low signal-to-noise ratio. \\
Third, as far as the H$\alpha$ intensity is concerned, Figures \ref{fig5-1} and \ref{fig5-2} provide a useful information on the proper line emission (lower than the L$\beta$ line by a factor eight). But in order to evaluate the feasibility of H$\alpha$ measurements in the corona and even possibly polarimetric ones, one definitely needs to take into account the continuum (Thomson) absorption and scattering. From Figures \ref{fig6-1}, \ref{fig6-2} and \ref{fig6-3}, one can compute the variation of the integrated intensity with the width of the integration band (or bandpass). \\
At 1.05 R$_{\odot}$, over a 2~\AA~bandpass, we find 8.6 erg~s$^{-1}$cm$^{-2}$sr$^{-1}$ for the streamer model and 0.16 erg~s$^{-1}$cm$^{-2}$sr$^{-1}$ at 1.9 R$_{\odot}$. For the quiet Sun, the values are still much lower (1.3 and 0.02 erg~s$^{-1}$cm$^{-2}$sr$^{-1}$, respectively, see Table \ref{tab1}). \\
Although these values are very low, the contrast of intensities between the streamer and the equatorial quiet-Sun is of the order of seven. This means that with the technique of background subtraction currently used in coronagraphic data, it is possible to access H$\alpha$ in the coronal extension of active regions, provided that the bandpass is equal or less than 2~\AA. \\
As far as coronal holes are concerned, the H$\alpha$ intensity (slightly dependent on the temperature) is 0.8 at 1.05 R$_{\odot}$ and $3.7\times 10^{-3}$ erg~s$^{-1}$cm$^{-2}$sr$^{-1}$ at 1.9 R$_{\odot}$. The contrast of the H$\alpha$ coronal hole (ratio to the quiet Sun) is 0.6 at 1.05 R$_{\odot}$ and 0.19 at 1.9 R$_{\odot}$. These numbers will actually be higher because of the LOS contamination; this leaves a small hope for detection in H$\alpha$ of out-of-the limb coronal holes.\\
The possibility of performing polarimetric measurements in H$\alpha$ has been discussed by \citet{2013SoPh..288..651K} who included the effect of instrumental stray light. Our computations show that the H$\alpha$ polarimetry in the corona could be envisaged above active regions with an instrumentation with a large aperture. However, the complexity of the line and its separation between polarizable and non-polarizable states (Dubau, 2015, private communication) make the interpretation complex, as noted by \citet{2012ApJ...749..136L} for the chromosphere. \\
Finally the H$_\alpha$ results are summarized in Table \ref{tab1}. \\
\begin{table}[H]
\begin{tabular}{lllllll}
\hline
Band- & Quiet & Quiet  & Coronal & Coronal  & Streamer & Streamer \\
pass & Sun & Sun & hole & hole &  &  \\
\AA & 1.05R$_{\odot}$ & 1.9R$_{\odot}$ & 1.05R$_{\odot}$ & 1.9R$_{\odot}$ & 1.05R$_{\odot}$ & 1.9R$_{\odot}$ \\
\hline
1 & $3.97~10^{-1}$ & $6.7~10^{-3}$ &  $2.31~10^{-1}$ & $1.06~10^{-3}$ & 2.91 & $4.72~10^{-2}$ \\
1.5 & $7.87~10^{-1}$ & $1.37~10^{-2}$ & $4.61~10^{-1}$ & $2.19~10^{-3}$ & 5.43 & $9.65~10^{-2}$ \\
2 & 1.29  & $2.3~10^{-2}$ &  $7.60~10^{-1}$ & $3.67~10^{-3}$ & 8.57 & $1.61~10^{-1}$ \\
3 & 2.42  & $4.4~10^{-2}$ & 1.43 & $7.01~10^{-3}$ & 15.5 & $3.07~10^{-1}$ \\
4 & 3.61 & $6.65~10^{-2}$ & 2.14 & $1.06~10^{-2}$ & 22.7 & $4.63~10^{-1}$ \\
5 & 4.86 & $9.03~10^{-2}$ & 2.89 & $1.44~10^{-2}$ & 30.1 & $6.27~10^{-1}$ \\
\hline
\end{tabular}
\caption{Integrated intensity [erg~s$^{-1}$cm$^{-2}$sr$^{-1}$] of the H$\alpha$ line as a function of the bandpass~\AA,~for the quiet-Sun, coronal-hole, and streamer models at positions 1.05 and 1.9 R$_{\odot}$ of the LOS.}\label{tab1}
\end{table}
\section{Conclusions}\label{section6}
Since we have the tools for computing exactly the ionization degree and the hydrogen--line emission in the corona, we envisage further improvements that will allow using more complex and realistic models.\\
First, we plan to include velocity fields which implies to compute a dilution factor taking into account the effect of velocities on the limb darkening or brightening of the incident radiation (the so-called ``Doppler dimming effect''\citep[see, \textit{e.g.},][]{1970SoPh...14..147H}. We also plan to take into account non-axisymetric illumination due to, {\it e.g.}, the proximity of an active region. Moreover, in order to compute exactly the electron population, we plan to include He in the ionization balance. Finally, we could also include a scattering dependent on the angle between incident and emergent radiations. We can also think of a modelling, where proton and electron temperatures are different \citep{1999A&A...347..676M}.\\
Second, we plan to use 2D or 3D coronal models (whether MHD or empirical) with full consistency between the various thermodynamic parameters, and complex geometries. \\
The combination of better tools for the computation of the ionization and the emission in the corona, along with realistic thermodynamic models of the corona, will allow for the interpretation of observations from future missions such as {\it Solar Orbiter}.

\begin{figure}[H]
\centering
\includegraphics[height=8cm]{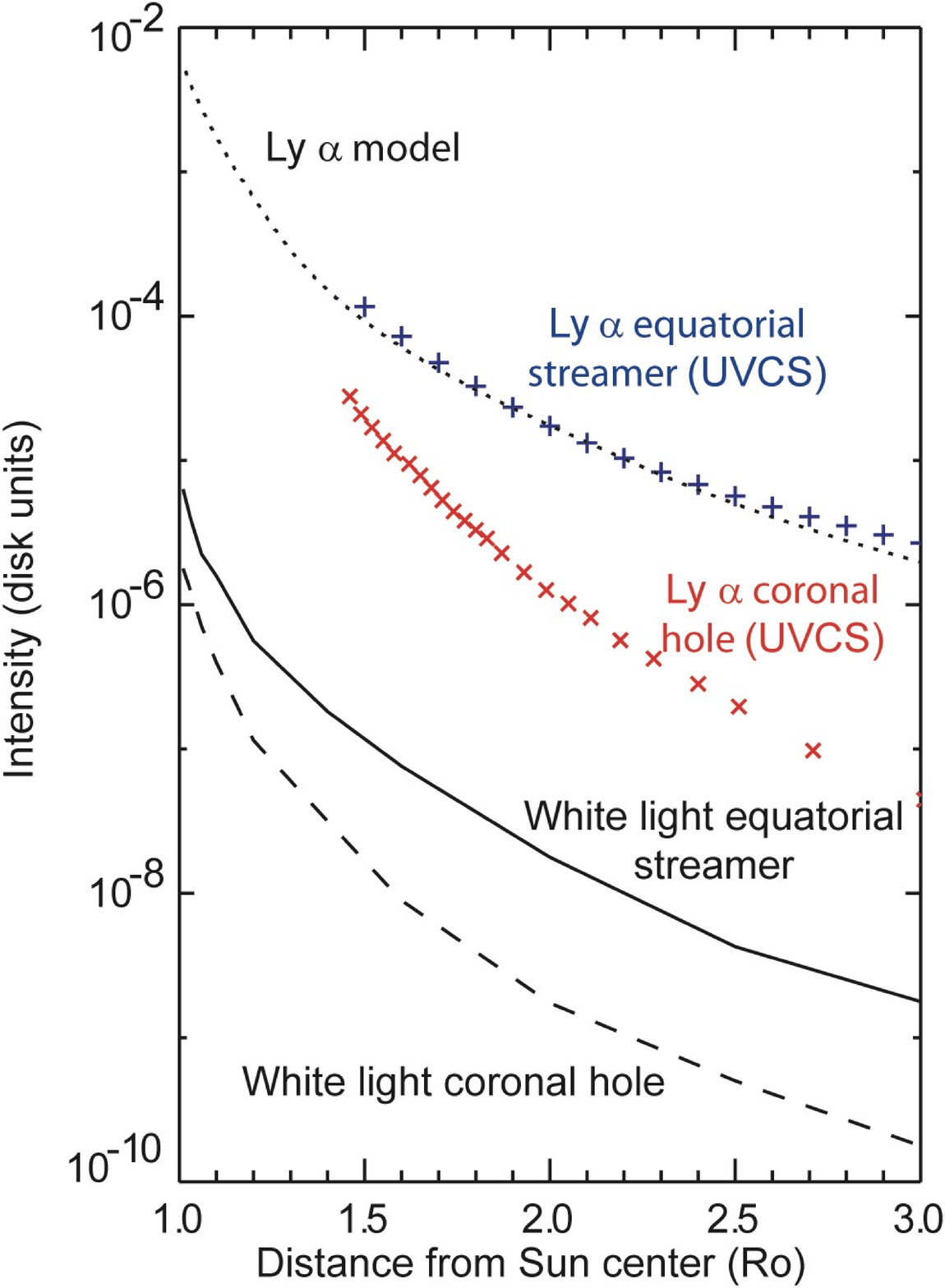}
\caption{Variation of the L$\alpha$ and white-light intensities (normalized to the disk-averaged intensities) with the distance to Sun center for two coronal regions: coronal hole and equatorial streamer. The L$\alpha$ data are from UVCS, the white-light data from S. Koutchmy (2006, private communication). Reproduced from \citet{2007AdSpR..40.1787V}}\label{fig8}
\end{figure}

\appendix
This appendix is focused upon a simple derivation of the H$\alpha$ coronal emission from (measured) L$\alpha$ emission.\\
We proceed in two steps. First, we only consider a three-level hydrogen atom which involves the L$\alpha$, L$\beta$, and H$\alpha$ lines and we exclude any polarization analysis. Then, we take into account the incident and scattered continua (essentially Thomson scattering) that we superimpose on the hydrogen emission in order to predict the actual emerging H$\alpha$ intensity.\\
We assume that the emission in these three lines results from only the process of resonance scattering. This means that level 3 is populated through two processes: L$\beta$ absorption from level 1 and H$\alpha$ absorption from level 2. We also assume as a first step, that the H$\alpha$ emission results only from the H$\alpha$ absorption and we neglect $3\rightarrow 2 \rightarrow 1$ cascade from L$\beta$ absorption. Consequently, we can compare the emissions in L$\alpha$ and H$\alpha$ which share the level 2.\\
Since we have some values of the L$\alpha$ intensity in the corona, we hope to be able to derive H$\alpha$ intensities. We can obtain a synthetic view of the variation of the L$\alpha$ intensity {\it versus} altitude in the corona for two types of structures: a streamer and a coronal hole (Figure \ref{fig8} taken from \citet{2007AdSpR..40.1787V}).\\

The H$\alpha$ absorbed and emitted intensity is:
\begin{equation}\label{eq1}
I^{H\alpha}(r_0) = B_{23}~\frac{h\nu^{H\alpha}}{4\pi}\int_{-\infty}^{+\infty}n_2(s)\int_{0}^{+\infty}\int_{\Omega}I_\mathrm{inc}^{H\alpha}(\nu,\omega)\frac{\mathrm{d}\omega}{4\pi}\phi^{H\alpha}(\nu,s)\mathrm{d}\nu~\mathrm{d}s
\end{equation}
The L$\alpha$ absorbed intensity is: 
\begin{equation}\label{eq2}
I^{L\alpha}(r_0) = B_{12}~\frac{h\nu^{L\alpha}}{4\pi}\int_{-\infty}^{+\infty}n_1(s)\int_{0}^{+\infty}\int_{\Omega}I_\mathrm{inc}^{L\alpha}(\nu,\omega)\frac{\mathrm{d}\omega}{4\pi}\phi^{L\alpha}(\nu,s)\mathrm{d}\nu~\mathrm{d}s
\end{equation}
The L$\alpha$ emitted intensity is: 
\begin{equation}\label{eq3}
I^{L\alpha}(r_0) = A_{21}~\frac{h\nu^{L\alpha}}{4\pi}\int_{-\infty}^{+\infty}n_2(s)\mathrm{d}s\int_{0}^{+\infty}\phi(\lambda)\mathrm{d}\lambda
\end{equation}
where $\phi(\nu,s)$ is the absorption profile at position $s$, where s is the distance of point $M$ along the ray taken from the impact point $P$ (at altitude $r_0$).\\
$I_\mathrm{inc}(\nu,\omega)$ (or $I_\mathrm{inc}(\lambda,\omega)$) is the incident radiation at position $M(s)$, depending on the frequency [$\nu$] (or wavelength [$\lambda$]) and solid angle [$\omega$].\\With the transformation $\displaystyle \mathrm{d}s=\frac{r\mathrm{d}r}{\sqrt{r^2-r_0^2}}$, Equations (\ref{eq1}), (\ref{eq2}) and (\ref{eq3}) become Volterra integral equations of the first kind. But the kernels are complex and include singularities. Consequently, we proceed with important simplifications where the quantity $\displaystyle \int I_\mathrm{inc}(\lambda,\omega)\frac{\mathrm{d}\omega}{4\pi}~\mathrm{d}\lambda~\phi(\lambda)$ $\displaystyle\Big(\mathrm{or}~\int I_\mathrm{inc}(\nu,\omega)~\frac{\mathrm{d}\omega}{4\pi}~\mathrm{d}\nu~\phi(\nu)\Big)$ is replaced in Equations (\ref{eq1}) and (\ref{eq2}) by \\ $<I_{\mathrm{emitted}}>\times w(r_0)\times \mathrm{FWHM}$, where $w(r_0)=0.5~\Big(1-\sqrt{1-\Big(\frac{R}{r_0}\Big)^2}\Big)$ is the dilution factor taken at $r_0$. Equation (\ref{eq1}) becomes: 
$$I^{H\alpha}(r_0) = B_{23}~\frac{h\nu^{H\alpha}}{4\pi}\int_{-\infty}^{+\infty}n_2(s)~\mathrm{ds}~<I_{\mathrm{emitted}}^{H\alpha}>~w(r_0)~\mathrm{FWHM}_{H\alpha}$$
where $<I_{\mathrm{emitted}}>$ is the chromospheric (incident) integrated emission that, when possible, we replace by the product of the half-intensity by the Full Width at Half Maximum (FWHM).\\
We have considered that $\displaystyle \phi^{H\alpha}(\nu,s)$ or $\displaystyle \phi^{H\alpha}(\lambda,s)$ depends (weakly) on temperature which is about $10^6$K. Then the Doppler width is $\displaystyle \Delta\lambda_D^{H\alpha} \approx 2.8$~\AA~and \\ $\mathrm{FWHM}(=2\sqrt{\log_e 2}~\Delta\lambda_D)=4.7$~\AA.~We also replace $\displaystyle \int n_2~\mathrm{d}s$ by $n_2(r_0)~\Delta s$,\\
where $\Delta s$ is the LOS distance over which $n_2$ is significant.\\
The incident radiation is taken from \citet{1961ZA.....53...37D} for H$\alpha$ and from \citet{1969SvA....12..599M} for the nearby continuum. We take an average value of 60 \% of the continuum over a FWHM of 4.7~\AA.~Equation (\ref{eq1}) becomes:
\begin{equation}\label{eq4}
I^{H\alpha}(r_0) = B_{23}~\frac{h\nu^{H\alpha}}{4\pi}~n_2(r_0)~\Delta s~<I_{\mathrm{emitted}}^{H\alpha}>~w(r_0)~\mathrm{FWHM}(H\alpha)
\end{equation}
Similarly, (\ref{eq2}) becomes: 
\begin{equation}
I^{L\alpha}(r_0) = B_{12}~\frac{h\nu^{L\alpha}}{4\pi}~n_1(r_0)~\Delta s~<I_{\mathrm{emitted}}^{L\alpha}>~w(r_0)~\mathrm{FWHM}(L\alpha)
\end{equation}
Note that we take the same $\Delta s$ along which the densities ($n_1$, $n_2$, ...) are significant. This is not different from computing the emission at position $P(r_0)$.\\
Then, Equation (\ref{eq3}) becomes:
\begin{equation}\label{eq6}
I^{L\alpha}(r_0) = A_{21}~\frac{h\nu^{L\alpha}}{4\pi}~n_2(r_0)~\Delta s~\mathrm{FWHM}(L\alpha)
\end{equation}
The ratio between Equations (\ref{eq4}) and (\ref{eq6}) gives:
\begin{equation}\label{eq7}
\frac{I^{H\alpha}(r_0)}{I^{L\alpha}(r_0)}=\frac{B_{23}}{A_{21}}~\frac{\nu^{H\alpha}}{\nu^{L\alpha}}~<I_{\mathrm{emitted}}^{H\alpha}>~w(r_0)~\frac{\mathrm{FWHM}(H\alpha)}{\mathrm{FWHM}(L\alpha)}
\end{equation}
The L$\alpha$ emission is taken from Figure \ref{fig8} at 1.05 and 1.9 R$_{\odot}$. \\
With the respective values of $w(r)$ (0.35 at 1.05 R$_{\odot}$ and 0.075 at 1.9 R$_{\odot}$), the ratio $\displaystyle \frac{\mathrm{FWHM}(H\alpha)}{\mathrm{FWHM}(L\alpha)}$ taken at $10^6$ K as 9.3, we find, for a 4.7~\AA~bandpass, 107.5 and 0.23 erg~s$^{-1}$cm$^{-2}$sr$^{-1}$ at 1.05 and 1.9 R$_{\odot}$ respectively. Table \ref{tab1} provides exact values which are about three times weaker at both altitudes. This agreement, which is well within an order of magnitude, is rather satisfactory when we take into account the many assumptions made in our analytical computation on one hand, and our use of coronal models that may well not represent reality, on the other hand. However, this computation has the advantage of providing a simple rule (Equation (\ref{eq7})) for deriving the H$\alpha$ intensity whenever and wherever the L$\alpha$ intensity is measured.


\begin{acks}
This paper is dedicated to Lola Salines who was murdered in Paris on 13 November 2015.
The authors thank Pierre Gouttebroze, Ira\"{\i}da Kim and Serge Koutchmy for useful comments. They deeply thank the referee for their comments which helped to improve the paper and clarify its aims.
\end{acks}

{\footnotesize
\paragraph*{Disclosure of Potential Conflict of Interest} The authors declare
that they have no conflicts of interest.
}

%
\newcommand\aj{Astronomical Journal}%
\newcommand\actaa{Acta Astronomica}%
\newcommand\araa{Annual Review of Astron and Astrophys}%
\newcommand\apj{Astrophys. J.}%
\newcommand\apjl{Astrophys. J. Lett.}%
\newcommand\apjs{Astrophys. J. Suppl.}%
\newcommand\apss{Astrophysics and Space Science}%
\newcommand\aap{Astron. Astrophys.}%
\newcommand\aapr{Astron. Astrophys. Reviews}%
\newcommand\aaps{Astron. Astrophys. Suppl.}%
\newcommand\azh{Astronomicheskii Zhurnal}%
\newcommand\baas{Bulletin of the AAS}%
\newcommand\caa{Chinese Astronomy and Astrophysics}%
\newcommand\cjaa{Chinese Journal of Astronomy and Astrophysics}%
\newcommand\icarus{Icarus}%
\newcommand\jcap{Journal of Cosmology and Astroparticle Physics}%
\newcommand\jrasc{Journal of the RAS of Canada}%
\newcommand\memras{Memoirs of the RAS}%
\newcommand\mnras{Monthly Notices of the RAS}%
\newcommand\na{New Astronomy}%
\newcommand\nar{New Astronomy Review}%
\newcommand\pra{Physical Review A: General Physics}%
\newcommand\prb{Physical Review B: Solid State}%
\newcommand\prc{Physical Review C}%
\newcommand\prd{Physical Review D}%
\newcommand\pre{Physical Review E}%
\newcommand\prl{Physical Review Letters}%
\newcommand\pasa{Publications of the Astron. Soc. of Australia}%
\newcommand\pasp{Publications of the ASP}%
\newcommand\pasj{Publications of the ASJ}%
\newcommand\rmxaa{Revista Mexicana de Astronomia y Astrofisica}%
\newcommand\qjras{Quarterly Journal of the RAS}%
\newcommand\skytel{Sky and Telescope}%
\newcommand\solphys{Solar Phys.}%
\newcommand\sovast{Soviet Astronomy}%
\newcommand\ssr{Space Sci. Rev.}%
\newcommand\zap{Zeitschrift fuer Astrophysik}%
\newcommand\nat{Nature}%
\newcommand\iaucirc{IAU Cirulars}%
\newcommand\aplett{Astrophysics Letters}%
\newcommand\apspr{Astrophysics Space Physics Research}%
\newcommand\bain{Bulletin Astronomical Institute of the Netherlands}%
\newcommand\fcp{Fundamental Cosmic Physics}%
\newcommand\gca{Geochimica Cosmochimica Acta}%
\newcommand\grl{Geophysics Research Letters}%
\newcommand\jcp{Journal of Chemical Physics}%
\newcommand\jgr{Journal of Geophysics Research}%
\newcommand\jqsrt{Journal of Quantitiative Spectroscopy and Radiative Transfer}%
\newcommand\memsai{Mem. Societa Astronomica Italiana}%
\newcommand\nphysa{Nuclear Physics A}%
\newcommand\physrep{Physics Reports}%
\newcommand\physscr{Physica Scripta}%
\newcommand\planss{Planetary Space Science}%
\newcommand\procspie{Proceedings of the SPIE}%

\bibliographystyle{spr-mp-sola}
\bibliography{bib.bib}

\end{article}
\end{document}